\documentclass[11pt]{article}
\usepackage{amsmath}
\usepackage{color}
\usepackage{fancyhdr}
\usepackage{verbatim}
\usepackage{amsfonts,longtable}
\usepackage{epsfig}
\usepackage{amsfonts}
\usepackage{color}

\def\R{\hbox{{\rm I}\kern-0.2em{\rm R}\kern0.2em}}
\def\a{\alpha}   
\def\b{\beta} \def\t{\theta}

\def\d{{\rm d}}

\def\de{\delta}

\def\g{\gamma}
\def\bn{\begin{equation}}
\def\en{\end{equation}}
\def\bny{\begin{eqnarray}}
\def\eny{\end{eqnarray}}
\def\be{\begin{eqnarray*}}
\def\ee{\end{eqnarray*}}
\def\bc{\begin{center}}
\def\ec{\end{center}}
 
\def\p{\partial}
\def\({\left(}
\def\){\right  )}
\def\[{\left[}
\def\]{\right]}
\def\bc{\begin{center}}
\def\ec{\end{center}}

\newtheorem{dfn}{Definition}[section]
\newtheorem{thm}{Theorem}[section]
\newtheorem{rem}{Remark}[section]
\newtheorem{pro}{Proposition}[section]

\newtheorem{cor}{Corollary}[section]
\newtheorem{lem}{Lemma}[section]
\newtheorem{exm}{Example}[section]

\def\bn{\begin{equation}}
\def\en{\end{equation}}
\def\bny{\begin{eqnarray}}
\def\eny{\end{eqnarray}}
\def\be{\begin{eqnarray*}}
\def\ee{\end{eqnarray*}}
\def\bdn{\begin{dfn}}
\def\edn{\end{dfn}}
\def\btm{\begin{thm}}
\def\etm{\end{thm}}
\def\bpf{\begin{proof}}
\def\epf{\end{proof}}
\def\bpn{\begin{pro}}
\def\epn{\end{pro}}
\def\brk{\begin{rem}}
\def\erk{\end{rem}}
\def\bcy{\begin{cor}}
\def\ecy{\end{cor}}
\def\blm{\begin{lem}}\def\elm{\end{lem}}
\def\bex{\begin{exm}}
\def\eex{\end{exm}}

 \def\R{{\hat R}}

\begin{document}

\bc {\bf Some geometric structures of wave equations on manifolds of `neutral signatures'
  }\ec
\medskip
\bc
Jean-Juste Bashingwa$^{}$ and  A. H. Kara$^{}\footnote{Department of Mathematics and Statistics, King Fahd University of Petroleum and Minerals, Dhahran, Saudi Arabia.}$
\\School of Mathematics, University of the Witwatersrand, Johannesburg, South Africa.\\
\ec
\begin{abstract}
In this paper, we study the symmetries and perform other geometric analyses of the wave equation on some spacetimes {with non diagonal metric $g_{ij}$ which are of neutral signatures}. Wave equations
on the standard Lorentzian manifolds have been done but not on the manifolds from metrics of neutral signatures.
\end{abstract}

\section{Introduction}

The symmetry classification problem for a number of wave equations has been studied in flat space  \cite{111,112,113,115} and non flat space (non-zero constant curvature) \cite{221,222}. In this work we pursue an investigation of symmetries of the wave equation on some spacetimes \textbf{with non diagonal metric $g_{ij}$ and neutral signatures}.  The corresponding ASD (anti self dual) manifolds have been studied in detail by a number of authors, for e.g., Fukaya \cite{f}, Dunajksi \cite{d}, Plebanski \cite{p} and Malykh et al \cite{m}, inter alia.

One of the more significant applications of Lie symmetry groups is to achieve a complete classification of symmetry reductions of partial differential equations.
The standard wave equation in $(3+1)$-dimensions has been extensively studied in the literature. A detailed symmetry analysis of this equation is discussed in \cite{Ibra,S10}.

\section{Wave equations on the ASD-Einstein manifold}

A detailed symmetry analysis of  ASD-Einstein manifolds has been done  in \cite{mine} .The metric on the ASD Ricci-flat  is locally given by
\bn
\d s^2=\d z \d y +\d t \d x -\left( -\frac{3z}{t^2}y +\frac{x}{t}\right)\d t^2 +\frac{2y}{t} \d t \d z \nonumber
\en

A Gordon type equation in a curved space or in curvilinear coordinates  is given by the expression
\bn
\square u=\frac{1}{\sqrt{|g|}}\left( g^{ab} u_{,b}\sqrt{|g|} \right)_{a}=k(u) \label{waveq}
\en

$g_{ab}$ being the metric of this space, $g=\parallel g_{ab} \parallel$ its determinant,  $g^{ab}$ the inverse of $g_{ab}$  , where $_{,b}=\p_b$ and $\square$ is the D'Alembertian, sometimes called the "box" operator.

Consequently, the Gordon type equation on the ASD manifofd takes the form

\bn
\frac{ xt- 3yz}{t^2}u_{xx}-\frac{2y}{t}u_{xy}+u_{x,t}+u_{yz}-k(u)=0.\label{wave1}
\en

\subsection{Lie symmetries}

The procedure for finding Lie point symmetries is well known  \cite{09} and thus will be presented without details. It turns out, from the symmetry study, that some special polynomial cases of $k(u)$ arise. Consequently, we study four cases for $k(u)$ in \eqref{wave1} given by

\begin{itemize}
\item[(i)] $k(u)=0$  (wave equation)
\item[(ii)] $k(u)=u$
\item[(iii)] $k(u)=u^3$
\item[(iv)] $k(u)=u^n,\quad n\neq 0,1,3$
\end{itemize}

\textbf{ Case (i)} $k(u)=0$. Following the symmetry criterion, we find that equation \eqref{wave1} in this case admits the following   fifteen Lie point symmetries,

\bn
\begin{array}{ll}
X_1=& u\p_u\\
X_2=&f_1(x,y,z,t)\p_u\\
X_3=&t\p_x \\
X_4=& 3t^{-1-\sqrt{7}}z\p_x + (2+\sqrt{7})t^{-\sqrt{7}}\p_y \\
X_5=& 3t^{-1+\sqrt{7}}z\p_x - (-2+\sqrt{7})t^{\sqrt{7}}\p_y \\
\end{array}\nonumber
\en
\bn
\begin{array}{ll}
X_6=&t\p_t+y\p_y\\
X_7=&2y \p_y+3t\p_t-x\p_x\\
X_8=&(-2+ \sqrt{7})t^{1-\sqrt{7}}y\p_x + t^{2-\sqrt{7}}\p_z\\
X_9=&(2+ \sqrt{7})t^{1+\sqrt{7}}y\p_x - t^{2+\sqrt{7}}\p_z\\
X_{10}=&t^{-\sqrt{7}}\left[(1-\sqrt{7})u y \p_u +\sqrt{7}  y^2 \p_y+ tx \p_z -ty\p_t + \frac{(-3+\sqrt{7})txy +3y^2z}{t} \p_x     \right ]\\
X_{11}=&t^{\sqrt{7}}\left[(-1-\sqrt{7})u y \p_u +\sqrt{7}  y^2 \p_y- tx \p_z +ty\p_t + \frac{(3+\sqrt{7})txy -3y^2z}{t} \p_x     \right ] \\
X_{12}=& 2t \p_t +y\p_y +z\p_z \\
X_{13}=& \frac{t^2 x^2 -txyz+3y^2z^2}{t^3}\p_x- \frac{u(tx +yz)}{t^2}\p_u +\frac{xz}{t}\p_z + \frac{y(tx+2yz)}{t^2}\p_y -\frac{yz}{t}\p_t \\
X_{14}=& t^{-2-\sqrt{7}} \left[ (1-\sqrt{7})uz\p_u +(-2+\sqrt{7})z^2\p_z -tz\p_t + (tx +2yz)\p_y + \frac{(-3+\sqrt{7})txz +3yz^2}{t}\p_x    \right]\\
X_{15}=& t^{-2+\sqrt{7}} \left[ (-1-\sqrt{7})uz\p_u +(2+\sqrt{7})z^2\p_z +tz\p_t - (tx +2yz)\p_y + \frac{(3+\sqrt{7})txz -3yz^2}{t}\p_x    \right]
\end{array} \label{wave2}
\en

where
$$ t^2f_{1,yz}+t^2f_{1,xt}-2tyf_{1,xy}+tx f_{1,xx}-3yzf_{1,xx}=0$$
\textbf{ Case (ii)} $k(u)=u$.  Equation \eqref{wave1} in this case admits    eight Lie point symmetries given by,

\bn
\begin{array}{ll}
X_1=&u\p_u\\
X_2=&f_1(x,y,z,t
) \p_u\\
X_3=&t\p_x\\
X_4=&-\frac{z(2f_2[t]+tf_2')}{t}+f_2[t]\p_y\\
X_5=&(-2+\sqrt{7})t^{1-\sqrt{7}}y\p_x+t^{2-\sqrt{7}}\p_z\\
X_6=&-(2+\sqrt{7})t^{1+\sqrt{7}}y\p_x+t^{2+\sqrt{7}}\p_z\\
X_7=&y\p_y-z\p_z\\
X_8=& t\p_t-x\p_x
\end{array}
\en
where $$-t^2f_1+2t^2f_{1,yz}+2t^2f_{1,xt}-4tyf_{1,xy}+2txf_{1,xx}-6yzf_{1,xx}=0$$ and $$7f_2 -tf_2'-t^2f_2''=0$$

\textbf{ Case (iii)} $k(u)=u^3$.  Equation \eqref{wave1} in this case admits    fourteen Lie point symmetries given by,

\bn
\begin{array}{ll}
X_1=& t\p_x \\
X_2=& -f_1(t)\p_y +z\left( \frac{2f_1(t)}{t}+f_1'\right)\p_x\\
X_3=& (-2+\sqrt{7})t^{1-\sqrt{7}}y\p_x + t^{2-\sqrt{7}}\p_z\\
X_4=& (2+\sqrt{7})t^{1+\sqrt{7}}y\p_x - t^{2+\sqrt{7}}\p_z\\
X_5=& -(-1+\sqrt{7})t^{-\sqrt{7}}uy\p_u +\sqrt{7}t^{-\sqrt{7}}y^2\p_y +t^{1-\sqrt{7}}x\p_z -t^{1-\sqrt{7}}y\p_t +t^{-1-\sqrt{7}}y((-3+\sqrt{7})xt +3yz)\p_x\\
X_6=& -(1+\sqrt{7})t^{\sqrt{7}}uy\p_u +\sqrt{7}t^{\sqrt{7}}y^2\p_y -t^{1+\sqrt{7}}x\p_z +t^{1+\sqrt{7}}y\p_t +t^{-1+\sqrt{7}}y((3+\sqrt{7})xt -3yz)\p_x\\
X_7=&y\p_y-z\p_z\\
X_8=& 2z\p_z +t\p_t-u\p_u+x\p_x\\
X_9=& 2x\p_x+2z\p_z -u\p_u\\
X_{10}=&t^{-2\sqrt{7}}\left[ (-1+\sqrt{7})u\p_u -(-2+\sqrt{7})z\p_z -\sqrt{7}y\p_y +\frac{tx+2(-7+2\sqrt{7})yz}{t}\p_x +t\p_t\right]\\
X_{11}=&t^{2\sqrt{7}}\left[ (1+\sqrt{7})u\p_u -(2+\sqrt{7})z\p_z -\sqrt{7}y\p_y +\frac{-tx+2(7+2\sqrt{7})yz}{t}\p_x -t\p_t\right]\\
X_{12}=&t^{-2-\sqrt{7}}\left[ -(-1+\sqrt{7})uz\p_u +(-2+\sqrt{7}){z^2}\p_z -tz\p_t + (tx +2yz)\p_y +\frac{z((-3+\sqrt{7})tx+3yz)}{t}\p_x   \right]\\
X_{13}=&t^{-2+\sqrt{7}}\left[ -(1+\sqrt{7})uz\p_u +(2+\sqrt{7}){z^2}\p_z +tz\p_t - (tx +2yz)\p_y +t^{-1}z((3+\sqrt{7})tx-3yz)\p_x   \right]\\
X_{14}=&\frac{t^2x^2-txyz+3y^2z^2}{t^3}\p_x +\frac{u(tx+yz)}{t^2}\p_u+\frac{xz}{t}\p_z +\frac{y(tx+2yz)}{t^2}\p_y -\frac{yz}{t}\p_t
\end{array}
\en
\textbf{ Case (iv)} $k(u)=u^n,\quad n\neq 0,1,3$.  Equation \eqref{wave1} in this case admits    eight Lie point symmetries given by,

\bn
\begin{array}{ll}
X_1=&t\p_x\\
X_2=&3t^{-1-\sqrt{7}}z\p_x +(2+\sqrt{7})t^{-\sqrt{7}}\p_y\\
X_3=&3t^{-1+\sqrt{7}}z\p_x -(-2+\sqrt{7})t^{\sqrt{7}}\p_y\\
X_4=&-\frac{u}{-1+n}\p_u+t\p_t+y\p_y\\
X_5=&-\frac{2u}{-1+n}\p_u+3t\p_t+2y\p_y -x\p_x\\
X_6=&-\frac{2u}{-1+n}\p_u+2t\p_t+y\p_y +z\p_z\\
X_7=& (-2+\sqrt{7})t^{1-\sqrt{7}}y\p_x +t^{2-\sqrt{7}}\p_z\\
X_8=& (2+\sqrt{7})t^{1+\sqrt{7}}y\p_x -t^{2+\sqrt{7}}\p_z
\end{array}
\en
\subsubsection{Symmetry reduction}

We demonstrate the reduction of the (1+3) dimensional wave equation \eqref{wave1}. The equation with four independent variables is reduced to a partial differential equation that has two independent variables. The reduced equation may then be analysed further using another Lie symmetry reduction or another appropriate method.

Since $[X,Y]=0$ with $X=t\p_t+y\p_y\quad Y=t\p_t-x\p_x$, where $X$ and $Y$ appear as Lie symmetries (also Noether symmetries as will be seen later) in all the above cases (even not explicitly but as linear combinations), we may begin reducing with $X$.

The characteristic equations are

$$\frac{\d x}{0}=\frac{\d y}{y}=\frac{\d z}{0}=\frac{\d t}{t}=\frac{\d u}{0}$$

Integrating yields $\a = \frac{y}{t}$ and \eqref{wave1} is reduced to
\bn
(2x -6\a z)u_{xx}-6\a u_{x \a}+2u_{z \a}=0 \label{wave4}
\en

with $u=u(x,\a,z)$.

If we then reduce \eqref{wave4} by $Y$, we obtain the transformation $\bar{Y}=-\a \p_\a-x\p_x$. We now have the characteristic equations,

$$\frac{\d x}{-x}=\frac{\d \a}{-\a}=\frac{\d z}{0}=\frac{\d u}{0}$$

By integrating, we obtain $\b=\frac{x}{\a}$ and \eqref{wave4} reduces to

\bn
(4\b-3z)u_{\b\b}+3u_{\b}-\b u_{\b z}=0\label{wave5}
\en

with $u=u(\b,z)$

Equation \eqref{wave5} may be further analysed or reduced using the underlying symmetries. The Lie point symmetries are given by

$$(f_1(\b,z)+ uf_2(z))\p_u +f_4(z)\p_z+f_3(\b,z)\p_\b$$

where
\bn\begin{array}{ll}
&3z f_3-3\b f_4-3z\b f_4'+4\b^2 f_4'+\b^2 f_{3,z}+3z\b f_{3,\b}-4\b^2 f_{3,\b}=0,\\
&-3f_{1,\b}+\b f_{1,\b z}+3z f_{1,\b \b}-4\b f_{1,\b \b}=0,\\
&12f_3-9f_4-3z\b f_2'+4\b^2f_2'+3\b f_{3,z}+9z f_{3,\b}-12\b f_{3,\b}+(3z\b -4\b^2)f_{3,\b z}+\\
&(9z^2 -24z \b +16\b^2)f_{3,\b\b}=0 \nonumber
\end{array}
\en
Take $f_1=0\quad f_2=1,\quad f_3=\b\quad f_4=z$, we have the characteristic equations

$$\frac{\d z}{z}=\frac{\d \b}{\b}=\frac{\d u}{u}$$

By integrating we obtain $\g=\frac{\b}{z}$ , $u=\frac{U}{z}, \quad U=U(\g)$ and \eqref{wave5} reduces to the second order ODE

$$(\g ^2+4\g -3)U_{\g \g}+(2\g +3)U_{\g}=0. $$

The solution is $$U(\gamma)=C_1+\exp \left[-\frac{1}{7}\arctan\text{h}\left( \frac{1}{7}(x+2)\sqrt{7}\right) \right]C_2$$ where $C_1,C_2$ are constant

\subsection{Noether symmetries}

Consider the wave equation \eqref{wave1}. Since it is variational, the corresponding Lagrangian is given by
\bn
L=\frac{tx-3yz}{t^2}u_x^2 -\frac{2y}{t} u_y
u_x+u_tu_x+u_zu_y -2h(u)\label{wave6}\en

where $h(u)=\int k(u) \d u$

\textbf{Case (i)} $h(u)=0$

Let

$$X=\xi(x,y,z,t,u)\p_x+\eta(x,y,z,t,u)\p_y+\gamma(x,y,z,t,u)\p_z
+\tau(x,y,z,t,u)\p_t+\phi(x,y,z,t,u)\p_u $$ be a Noether point operator 
with gauge vector $f_i(i=1,2,3,4)$ dependent on $(t,x,y,z,u)$. This becomes, for Lagrangian given by \eqref{wave6},

$$XL+L[D_x \xi+ D_y \eta +D_z \gamma +D_t \tau ]=D_x f_1 +D_y f_2+D_z f_3+D_t f_4$$
Separation by derivatives of $u$ yields the following overdetermined system

\bn \begin{array}{ll}
u_x^3 &:  \xi_u \\
u_x^2 u_y&:\eta_u\\
u_x^2 u_t&:\tau_u\\
u_xu_tu_z&:\gamma_u\\
u_x u_y&:  -2 yt^{2}\tau_t -4{t}^{2}y\phi_u -t \left( 2xt-6
yz \right) \eta_x - \eta_t {t}^{3}-2{t}^{2}y\gamma_z -{t}^{3}\xi_z +2yt\tau  -2
{t}^{2}\eta  \\
u_xu_z&: -t \left( 2xt-6yz \right) \gamma_x - \gamma_t {t}^{
3}+2{t}^{2}y\gamma_y -{t}^{3}\xi_y \\
 u_x u_t&: 2{t}^{3}
\phi_u -t
 \left( 2xt-6yz \right) \tau_x +2{t}^{2}y\tau_y +{t}^{3}\gamma_z +{t}^{3}\eta_y  \\
 u_yu_z&: 2 \gamma_x {t}^{2}y+2{t}^{3}\phi_u + \tau_t  {t}^{3}+
 \xi_x {t}^{3}\\
u_yu_t &: -
 \eta_x {t}^{3}+2\tau_x {t}^{2}y- \tau_z {t}^{3} \\
 u_zu_t&: - \gamma_x {t}^{3}- \tau_y
{t}^{3} \\
u_x^2&:
   \left( xt-3
yz \right) (2t\phi_u+ t\tau_t+t\gamma_z+t\eta_y-t\xi_x)   +2{t}^{2}y\xi_y -{t}^{3}\xi_t - \left( xt
-6yz \right) \tau  + \left( -3z\eta
  +t\xi  -3y\gamma \right) t \\

u_y^2 &: 2y\eta_x - \eta_z {t} \\
u_z^2 &- \gamma_y\\
u_t^2 &:\tau_x\\
u_z&\phi_y - f_{3,u}\\
 u_x &: t \left( 2xt-6yz \right) \phi_x + \phi_t {t}^{3}- f_{1,u}t^3
-2{t}^{2}y\phi_y   \\
u_t&:
\phi_x - f_{4,u}  \\
u_y&:
 -2y\phi_x + \phi_z {t}- f_{2,u} {t}
 \\
1&:f_{1,x}+f_{2,y}+f_{3,z}+f_{4,z}

\end{array}\en

The detailed calculations lead to
\bn
\begin{array}{ll}
X_1=& x\p_x +y\p_y-\frac{u}{2}\p_u\\
X_{2}=&-x\p_x+t\p_t\\
X_{3}=& x\p_x +z\p_z -\frac{u}{2}\p_u\\
X_4=& t^{\sqrt{7}}\left[-\frac{z(2+\sqrt{7})}{t}\p_x +\p_y  \right]\\
X_5=& \frac{1}{t^{\sqrt{7}}}\left[\frac{z(-2+\sqrt{7})}{t}\p_x +\p_y  \right]\\
X_{6}=&-t^{1+\sqrt{7}}y(2+\sqrt{7})\p_x + t^{2+\sqrt{7}}\p_z\\
X_{7}=&yt^{1-\sqrt{7}}(-2+\sqrt{7})\p_x +t^{2-\sqrt{7}}\p_z\\

X_{8}=&t\p_x\\
X_9=& \frac{t^2x^2-txyz+3y^2z^2}{2t^3}\p_x +\frac{y(tx+2yz)}{2t^2}\p_y + \frac{xz}{2t}\p_z -\frac{yz}{2t}\p_t - \frac{u(tx+yz)}{2t^2}\p_u\\
X_{10}=&t^{\sqrt{7}}\left[-\frac{z(\sqrt{7}tx +3xt-3yz)}{2t^3}\p_x + \frac{tx+2yz}{2t^2}\p_y -\frac{z^2(2+\sqrt{7})}{2t^2}\p_z-\frac{z}{2t}\p_t +\frac{zu(1+\sqrt{7})}{2t^2}\p_u     \right]\\
X_{11} = &\frac{1}{t^{\sqrt{7}}}\left[\frac{z(\sqrt{7}tx -3xt+3yz)}{2t^3}\p_x + \frac{tx+2yz}{2t^2}\p_y +\frac{z^2(-2+\sqrt{7})}{2t^2}\p_z-\frac{z}{2t}\p_t -\frac{zu(-1+\sqrt{7})}{2t^2}\p_u     \right]\\
X_{12}=&t^{\sqrt{7}}\left[\frac{y(3\sqrt{7}tx -3 \sqrt{7}yz+7xt)}{14t}\p_x  +\frac{y^2}{2}\p_y -\frac{\sqrt{7}tx}{14}\p_z +\frac{\sqrt{7}ty}{14}\p_t -\frac{yu(\sqrt{7}+7)}{14}\p_u     \right]\\
X_{13}=&\frac{1}{t^{\sqrt{7}}}\left[\frac{-y(3\sqrt{7}tx -3 \sqrt{7}yz-7xt)}{14t}\p_x  +\frac{y^2}{2}\p_y +\frac{\sqrt{7}tx}{14}\p_z -\frac{\sqrt{7}ty}{14}\p_t +\frac{yu(\sqrt{7}-7)}{14}\p_u     \right]\\

X_{14}=&t^{2\sqrt{7}}\left[ \frac{\sqrt{7}tx -14\sqrt{7}yz -28yz}{7t}\p_x +y\p_y +\frac{z(7+2\sqrt{7})}{7}\p_z +\frac{\sqrt{7}t}{7}\p_t -\frac{7+\sqrt{7}}{7}u\p_u            \right]\\
X_{15}=&\frac{1}{t^{2\sqrt{7}}}\left[ -\frac{\sqrt{7}tx -14\sqrt{7}yz +28yz}{7t}\p_x +y\p_y -\frac{z(-7+2\sqrt{7})}{7}\p_z -\frac{\sqrt{7}t}{7}\p_t +\frac{-7+\sqrt{7}}{7}u\p_u            \right]\label{wave8}

\end{array}
\en

The conserved forms corresponding to each Noether symmetry (\cite{N11}) is a three form $\omega$ such that the four form $D\omega$ vanishes. Thus

$$\omega=\Phi^x \d y\wedge \d z \wedge \d t-\Phi^y \d x\wedge \d z \wedge \d t +\Phi^z \d x\wedge \d y  \wedge \d t -\Phi^t \d x\wedge \d y \wedge \d z$$ so that $$D_x\Phi^x+D_y\Phi^y+D_z\Phi^z+D_t\Phi^t=0$$

We list all conserved flux for one case, and only the conserved densities for the remaining symmetries.
\bn
\begin{array}{ll}
\Phi_1^x=& \frac{1}{t^2}(2t^2xu_yu_z-2t^2yu_tu_y-2tx^2 u_x^2-4txyu_xu_y +4ty^2 u_y^2+6xyzu_x^2+12y^2zu_xu_y-t^2uu_t-\\& 2tuxu_x+2tyuu_y+6uyzu_x)\\
\Phi_1^y=& \frac{1}{t^2}\left(  2t^2xu_xu_z-2t^2 yu_tu_x -6txy u_x^2 +6y^2z u_x^2 +t^2 uu_z-2tuyu_x\right)\\
\Phi_1^z=& u_y(2xu_x+2yu_y+u)\\
\Phi_1^t=& u_x(2xu_x+2yu_y+u)\\
\Phi_2^t=& \frac{1}{t}(t^2 u_yu_z+2txu_x^2-2tyu_xu_y-3yz u_x^2)\\
\Phi_3^t=& -\frac{u_x}{2}(2 x u_x +2zu_z +u)\\
\Phi_4^t= & u_x(u_x t^{-1+\sqrt{7}} z \sqrt{7} +2u_x t^{-1+\sqrt{7}} -u_y t^{\sqrt{7}})\\
\Phi_5^t= &- u_x(u_x t^{-1-\sqrt{7}} z \sqrt{7} -2u_x t^{-1-\sqrt{7}} +u_y t^{-\sqrt{7}})\\
\Phi_6^t=& u_{{x}} ( u_{{x}}{t}^{1+ \sqrt{7}}y \sqrt{7}+2u_{{x}}{t}^{1+ \sqrt{7}}y-u_{{z}}{t}^{2+ \sqrt{7}} ) \\
\Phi_7^t= &-u_{x} ( u_{{x}}{t}^{1- \sqrt{7}}y \sqrt{7}-2u_{x}{t}^{1- \sqrt{7}}y+u_{z}{t}^{2- \sqrt{7}})\\
\Phi_8^t=&-{u_{{x}}}^{2}t\\
\Phi_9^t=&-\frac{1}{{2{t}^{2}}} (t{x}^{2}{u_{{x}}}^{2}+txyu_{{x}}u_{{y}}+txzu_{{x}}u_{{z}}
+tyzu_{{y}}u_{{z}}+tuxu_{{x}}+uyzu_{{x}})+\frac{u}{2t}\\

\Phi_{10}^t=& \frac{1}{-2}{t}^{-1+ \sqrt{7}}zu_{{y}}u_{{z}}+{t}^{-2+ \sqrt{7}}xz{u_{{x}}}^{2}+\frac{1}{2} \sqrt{7}{t}^{-2+ \sqrt{7}}xz{u_{{x}}}^{2}+\frac{1}{2}\sqrt{7}{t}^{-2+ \sqrt{7}}{z}^{2}u_{{x}}u_{{z}}+\frac{1}{2}\sqrt{7}{t}^{-2+ \sqrt{7}}zuu_{{x}}+\\
&{t}^{-2+ \sqrt{7}}{z}^{2}u_{{x}}u_{{z}}-\frac{1}{2}{t}^{-1+ \sqrt{7}}xu_{{x}}u_{{y}}+\frac{1}{2}{t}^{-2+ \sqrt{7}}zuu_{{x}} \\

\Phi_{11}^t=&\frac{1}{2}\left(-{t}^{-1- \sqrt{7}}zu_{{y}}u_{{z}}+2{t}^{-2- \sqrt{7}}xz{u_{{x}}}^{2}-{t}^{-2- \sqrt{7}} \sqrt{7}xz{u_{{x}}}^{2}-{t}^{-2- \sqrt{7}} \sqrt{7}{z}^{2}u_{{x}}u_{{z}}-\right.\\
&\left.{t}^{-2- \sqrt{7}}\sqrt{7}zuu_{{x}}+2{t}^{-2-\sqrt{7}}{z}^{2}u_{{x}}u_{{z}}+{t}^{-2- \sqrt{7}}zuu_{{x}}-{t}^{-1- \sqrt{7}}xu_{{x}}u_{{y}}\right)  \\

\Phi_{12}^t=&\frac{2}{7} \left( \sqrt{7}{t}^{1+ \sqrt{7}}yu_{{y}}u_{{z}}-2{t}^{ \sqrt{7}} \sqrt{7}xy{u_{{x}}}^{2}-2\sqrt{7}{t}^{ \sqrt{7}}{y}^{2}u_{{x}}u_{{y}}+{t}^{1+ \sqrt{7}} \sqrt{7}xu_{{x}}u_{{z}}-\,{t}^{ \sqrt{7}} \sqrt{7}yuu_{{x}}\right.\\& \left. -14{t}^{ \sqrt{7}}xy{u_{{x}}}^{2}-14{t}^{ \sqrt{7}}{y}^{2}u_{{x}}u_{{y}}-14{t}^{ \sqrt{7}}yuu_{{x}}\right) \\

\Phi_{13}^t =&\frac{1}{14}\left(-{t}^{1- \sqrt{7}} \sqrt{7}yu_{{y}}u_{{z}}+2{t}^{- \sqrt{7}} \sqrt{7}xy{u_{{x}}}^{2}+2\sqrt{7}{t}^{- \sqrt{7}}{y}^{2}u_{{x}}u_{{y}}+{t}^{- \sqrt{7}} \sqrt{7}yuu_{{x}}-7{t}^{- \sqrt{7}}xy{u_{{x}}}^{2}-\right. \\ & \left. 7{t}^{- \sqrt{7}}{y}^{2}u_{{x}}u_{{y}}-{t}^{1- \sqrt{7}} \sqrt{7}xu_{{x}}u_{{z}}-7{t}^{- \sqrt{7}}yuu_{{x}}\right) \\

\Phi_{14}^t =&\frac{1}{7}\left(\sqrt{7}{t}^{2\sqrt{7}+1}u_{{y}}u_{{z}}-2\sqrt{7}{t}^{2\sqrt{7}}yu_{{x}}u_{{y}}+11{t}^{2\sqrt{7}-1} \sqrt{7}yz{u_{{x}}}^{2}-2{t}^{2\sqrt{7}} \sqrt{7}zu_{{x}}u_{{z}}+28{t}^{2 \sqrt{7}-1}yz{u_{{x}}}^{2}-\right.\\&\left.7{t}^{2\sqrt{7}} \sqrt{7}uu_{{x}}-7{t}^{2 \sqrt{7}}yu_{{x}}u_{{y}}-7{t}^{2\sqrt{7}}zu_{{x}}u_{{z}}-7{t}^{2 \sqrt{7}}uu_{{x}}\right) \\

\Phi_{15}^t=& -\frac{1}{7}\left(\sqrt{7}{t}^{1-2\sqrt{7}}u_{{y}}u_{{z}}-2\sqrt{7}{t}^{-2\sqrt{7}}yu_{{x}}u_{{y}}+11{t}^{-2\sqrt{7}-1} \sqrt{7}yz{u_{{x}}}^{2}-2{t}^{-2\sqrt{7}} \sqrt{7}zu_{{x}}u_{{z}}-28{t}^{-2 \sqrt{7}-1}yz{u_{{x}}}^{2}\right.\\&\left. -{t}^{-2\sqrt{7}} \sqrt{7}u_{{x}}+7{t}^{-2 \sqrt{7}}yu_{{x}}u_{{y}}+7{t}^{2\sqrt{7}}zu_{{x}}u_{{z}}-7{t}^{-2 \sqrt{7}}u_{{x}}\right) \\
\end{array}
\en

\textbf{Case (ii)} $h(u)=u^n, \quad n \neq 0,1 $

It can be shown that a 7-dimensional algebra of point symmetry generators with basis (Noether symmetries) is given by

\bn \begin{array}{ll}
&X_1=-x\p_x+t\p_t, \qquad X_2=-y\p_y + z\p_z\qquad X_3=-(2+\sqrt{7})t^{1+\sqrt{7}}y\p_x+t^{2+\sqrt{7}}\p_z \qquad \\
&X_4=(-2+\sqrt{7})t^{1-\sqrt{7}}y\p_x+t^{2-\sqrt{7}}\p_z \qquad X_5=-(2+\sqrt{7})t^{-1+\sqrt{7}}z\p_x+t^{\sqrt{7}}\p_y \quad\\& X_6=(-2+\sqrt{7})t^{-1-\sqrt{7}}z\p_x+t^{-\sqrt{7}}\p_y \qquad X_7=t\p_x
\end{array}
\en
The corresponding conserved quantities are

\bn
\begin{array}{ll}
\Phi_1^t=& {\frac {-u_{{y}}u_{{z}}{t}^{2}-2\,tx{u_{{x}}}^{2}+2\,yu_{{x}}u_{{y}}t+3\,yz{u_{{x}}}^{2}+{u}^{n}{t}^{2}}{t}}\\
\Phi_1^z=&u_{{y}} \left( tu_{{t}}-xu_{{x}} \right) \\
\Phi_1^y=&-{\frac { \left( tu_{{z}}-2\,yu_{{x}} \right)  \left( tu_{{t}}-xu_{{x}} \right) }{t}} \\

\Phi_1^x=& -{\frac {-{t}^{3}{u_{{t}}}^{2}
\mbox{}-2\,{t}^{2}xu_{{x}}u_{{t}}-{t}^{2}xu_{{y}}u_{{z}}+2\,{t}^{2}yu_{{y}}u_{{t}}+t{x}^{2}{u_{{x}}}^{2}+6\,tyzu_{{x}}u_{{t}}-3\,xyz{u_{{x}}}^{2}+{u}^{n}{t}^{2}x
\mbox{}}{{t}^{2}}}\\
\Phi_2^t=& u_{{x}} \left( yu_{{y}}-zu_{{z}} \right)\\
\Phi_2^z=& {\frac {-{t}^{2}y{u_{{y}}}^{2}-{t}^{2}zu_{{x}}u_{{t}}-txz{u_{{x}}}^{2}+2tyzu_{{x}}u_{{y}}+3y{z}^{2}{u_{{x}}}^{2}\mbox{}+{u}^{n}{t}^{2}z}{{t}^{2}}} \\

\Phi_2^y=&{\frac {-{t}^{2}yu_{{x}}u_{{t}}-{t}^{2}z{u_{{z}}}^{2}-txy{u_{{x}}}^{2}+2tyzu_{{x}}u_{{z}}+3{y}^{2}z{u_{{x}}}^{2}+{u}^{n}{t}^{2}y}{{t}^{2}}} \\
\Phi_2^x=& -{\frac { \left( {t}^{2}u_{{t}}+2txu_{{x}}-2tyu_{{y}}-6yzu_{{x}} \right) \left( yu_{{y}}-zu_{{z}} \right) }{{t}^{2}}}\\
\Phi_3^t=& -u_{{x}} \left( -u_{{x}}{t}^{1+ \sqrt{7}}y \sqrt{7}-2u_{{x}}{t}^{1+ \sqrt{7}}y+u_{{z}}{t}^{2+\sqrt{7}} \right)\\
\Phi_3^z=& (3{t}^{ \sqrt{7}}yz{u_{{x}}}^{2}- \sqrt{7}{t}^{1+ \sqrt{7}}yu_{{x}}u_{{y}}-{t}^{1+ \sqrt{7}}x{u_{{x}}}^{2}-{t}^{2+ \sqrt{7}}u_{{x}}u_{{t}}+{t}^{2+ \sqrt{7}}{u}^{n})\\
\Phi_3^y=&-{\frac { \left( tu_{{z}}-2yu_{{x}} \right) \left( -u_{{x}}{t}^{1+ \sqrt{7}}y \sqrt{7}-2u_{{x}}{t}^{1+ \sqrt{7}}y+u_{{z}}{t}^{2+ \sqrt{7}} \right) }{t}} \\
\Phi_3^x=& \frac {1}{t^2}\left(3\sqrt{7}{t}^{1+ \sqrt{7}}{y}^{2}z{u_{{x}}}^{2}+6{t}^{1+ \sqrt{7}}{y}^{2}z{u_{{x}}}^{2}-{t}^{2+ \sqrt{7}} \sqrt{7}xy{u_{{x}}}^{2}-2{t}^{2+ \sqrt{7}}xy{u_{{x}}}^{2}+ \sqrt{7}{t}^{3+ \sqrt{7}}yu_{{y}}u_{{z}}-\right. \\&\left. \sqrt{7}{u}^{n}{t}^{3+ \sqrt{7}}y+{t}^{4+ \sqrt{7}}u_{{z}}u_{{t}}+2{t}^{3+ \sqrt{7}}xu_{{x}}u_{{z}}-6{t}^{2+ \sqrt{7}}yzu_{{x}}u_{{z}}-2{u}^{n}{t}^{3+ \sqrt{7}}y\right)\\

\end{array}\nonumber
\en
\bn
\begin{array}{ll}
\Phi_4^t=&-u_{{x}} \left( u_{{x}}{t}^{1- \sqrt{7}}y \sqrt{7}-2u_{{x}}{t}^{1- \sqrt{7}}y+u_{{z}}{t}^{2- \sqrt{7}} \right)  \\
\Phi_4^z=& ({t}^{1- \sqrt{7}} \sqrt{7}yu_{{x}}u_{{y}}+3{t}^{- \sqrt{7}}yz{u_{{x}}}^{2}-{t}^{1-\sqrt{7}}x{u_{{x}}}^{2}-{t}^{2- \sqrt{7}}u_{{x}}u_{{t}}+{t}^{2- \sqrt{7}}{u}^{n})\\
\Phi_4^y=& -{\frac { \left( tu_{{z}}-2yu_{{x}} \right)  \left( u_{{x}}{t}^{1- \sqrt{7}}y \sqrt{7}-2u_{{x}}{t}^{1- \sqrt{7}}y+u_{{z}}{t}^{2- \sqrt{7}} \right) }{t}}\\
\Phi_4^x=& -\frac {1}{t^2}\left(3{t}^{1- \sqrt{7}} \sqrt{7}{y}^{2}z{u_{{x}}}^{2}-6{t}^{1- \sqrt{7}}{y}^{2}z{u_{{x}}}^{2}-{t}^{2- \sqrt{7}} \sqrt{7}xy{u_{{x}}}^{2}+{t}^{3- \sqrt{7}} \sqrt{7}yu_{{y}}u_{{z}}+2{t}^{2- \sqrt{7}}xy{u_{{x}}}^{2}-\right. \\& \left.{t}^{3- \sqrt{7}} \sqrt{7}{u}^{n}y+2{t}^{3- \sqrt{7}}{u}^{n}y-{t}^{4- \sqrt{7}}u_{{z}}u_{{t}}-2{t}^{3- \sqrt{7}}xu_{{x}}u_{{z}}+6{t}^{2- \sqrt{7}}yzu_{{x}}u_{{z}}\right)\\
\Phi_5^t=& u_{{x}} \left( u_{{x}}{t}^{-1+ \sqrt{7}}z \sqrt{7}+2u_{{x}}{t}^{-1+ \sqrt{7}}z-u_{{y}}{t}^{ \sqrt{7}}\right)\\
\Phi_5^z=& -u_{{y}} \left( u_{{x}}{t}^{-1+ \sqrt{7}}z \sqrt{7}+2u_{{x}}{t}^{-1+ \sqrt{7}}z-u_{{y}}{t}^{\sqrt{7}}\right)\\
\Phi_5^y=& -{\frac {-{t}^{ \sqrt{7}} \sqrt{7}zu_{{x}}u_{{z}}+2{t}^{-1+ \sqrt{7}} \sqrt{7}yz{u_{{x}}}^{2}+7{t}^{-1+ \sqrt{7}}yz{u_{{x}}}^{2}-{t}^{ \sqrt{7}}x{u_{{x}}}^{2}-2{t}^{ \sqrt{7}}zu_{{x}}u_{{z}}-{t}^{1+ \sqrt{7}}u_{{x}}u_{{t}}+{t}^{1+ \sqrt{7}}{u}^{n}}{t}}\\
\Phi_5^x=& \frac {1}{t^2}\left[3{t}^{-1+ \sqrt{7}} \sqrt{7}y{z}^{2}{u_{{x}}}^{2}- \sqrt{7}{t}^{ \sqrt{7}}xz{u_{{x}}}^{2}+6{t}^{-1+ \sqrt{7}}y{z}^{2}{u_{{x}}}^{2}+{t}^{1+ \sqrt{7}} \sqrt{7}zu_{{y}}u_{{z}}-2{t}^{ \sqrt{7}}xz{u_{{x}}}^{2}-\right.\\&\left.6{t}^{ \sqrt{7}}yzu_{{x}}u_{{y}}-{t}^{1+ \sqrt{7}} \sqrt{7}{u}^{n}z+2{t}^{1+ \sqrt{7}}zu_{{y}}u_{{z}}-2{t}^{1+ \sqrt{7}}{u}^{n}z+{t}^{2+ \sqrt{7}}u_{{y}}u_{{t}}+2{t}^{1+ \sqrt{7}}xu_{{x}}u_{{y}}\right.\\&\left.-2{t}^{1+ \sqrt{7}}y{u_{{y}}}^{2}\right]\\
\Phi_6^t=& -u_{{x}} \left( u_{{x}}{t}^{-1- \sqrt{7}}z \sqrt{7}-2u_{{x}}{t}^{-1- \sqrt{7}}z+u_{{y}}{t}^{- \sqrt{7}}\right)\\
\Phi_6^z=&u_{{y}} \left( u_{{x}}{t}^{-1- \sqrt{7}}z \sqrt{7}-2u_{{x}}{t}^{-1- \sqrt{7}}z+u_{{y}}{t}^{- \sqrt{7}}\right) \\
\Phi_6^y=& {\frac {-7{t}^{-1- \sqrt{7}}yz{u_{{x}}}^{2}-{t}^{- \sqrt{7}} \sqrt{7}zu_{{x}}u_{{z}}+2{t}^{-1- \sqrt{7}} \sqrt{7}yz{u_{{x}}}^{2}+{t}^{- \sqrt{7}}x{u_{{x}}}^{2}+2{t}^{- \sqrt{7}}zu_{{x}}u_{{z}}+{t}^{1- \sqrt{7}}u_{{x}}u_{{t}}-{t}^{1- \sqrt{7}}{u}^{n}}{t}}\\
\end{array}\en

\bn\begin{array}{ll}

\Phi_6^x=& -\frac {1}{t^2}\left(3{t}^{-1- \sqrt{7}} \sqrt{7}y{z}^{2}{u_{{x}}}^{2}-{t}^{- \sqrt{7}} \sqrt{7}xz{u_{{x}}}^{2}-6{t}^{-1- \sqrt{7}}y{z}^{2}{u_{{x}}}^{2}+2{t}^{- \sqrt{7}}xz{u_{{x}}}^{2}+6{t}^{- \sqrt{7}}yzu_{{x}}u_{{y}}+\right.\\&  {t}^{1- \sqrt{7}} \sqrt{7}zu_{{y}}u_{{z}}-{t}^{1- \sqrt{7}} \sqrt{7}{u}^{n}z-2{t}^{1- \sqrt{7}}zu_{{y}}u_{{z}}+2{t}^{1- \sqrt{7}}{u}^{n}z-{t}^{2- \sqrt{7}}u_{{y}}u_{{t}}-2{t}^{1- \sqrt{7}}xu_{{x}}u_{{y}}+\\&\left. 2{t}^{1- \sqrt{7}}y{u_{{y}}}^{2}\right)\\
\Phi_7^t=&-{u_{{x}}}^{2}t \\
\Phi_7^z=&+u_{{y}}tu_{{x}} \\
\Phi_7^y=& - \left( tu_{{z}}-2yu_{{x}} \right) u_{{x}}\\
\Phi_7^x=& {\frac {-u_{{y}}u_{{z}}{t}^{2}+tx{u_{{x}}}^{2}-3yz{u_{{x}}}^{2}+{u}^{n}{t}^{2}}{t}}\\

\end{array}
\en

\subsection{Variational symmetries, multipliers approach}

Consider the wave equation \eqref{wave1} in ASD-Einstein spacetime with dependent variable $u=u(x,y,t)$, i.e., we have taken out the spatial variable $z$ from the original equation because the calculations with $z$ are extremely cumbersome producing no final outcomes. We consider the multiplier method for \eqref{wave1}. Let us choose $k(u)=-u$, we have

$$\frac{\delta}{\delta u}\left[\mathcal{Q}\left(\frac{ x }{t}u_{xx}-\frac{2y}{t}u_{xy}+u_{x,t}+u \right)\right]=0$$ where $\mathcal{Q}=\mathcal{Q}(x,y,u,u_x,u_y,u_{x,x},u_{y,y},u_{x,y},
u_{x,x,x},u_{x,x,y})$. Even though not pursued here, the calculation may include $t$ and derivatives of $u$ with respect to $t$. Then

$$\frac{\delta}{\delta u}\left[\mathcal{Q}\left(\frac{ x }{t}u_{xx}-\frac{2y}{t}u_{xy}+u_{x,t}+u \right)\right]=D_t \Phi^t +D_y\Phi^y+D_x\Phi^x$$

where $(\Phi^x,\Phi^y,\Phi^t)$ is the conserved flow. We obtain the set of multiplier $\mathcal{Q}_i$ together with the conserved densities $\Phi_i^t$, namely,

\bn
\begin{array}{ll}
\mathcal{Q}_1 =&\frac{{u}_x}{\sqrt{y}}\\
%
 \Phi_1^t=& \frac{{u}_{{x}}{}^2-{u} {u}_{{ }{x}{x}}}{4 \sqrt{y}} \\

\mathcal{Q}_2 =&\frac{1}{ty}(t{u}_x+x{u}_{xxx}+{u}_{xx})\\
 \Phi^t_2=& \frac{1}{4 t y}[t {u}_{{ }{x}}{}^2+\left({u}_{{ }{x}{x}}+x {u}_{{ }{x}{x}{x}}\right) {u}_{{x}}-u \left(t {u}_{{x}{x}}+2 {u}_{{x}{x}{x}}+x {u}_{{x}{x}{x}{x}}\right)] \\

\mathcal{Q}_3 =&y{u}_y+\frac{1}{2}{u}\\
  \Phi^3_t = & \frac{1}{4} y ({u}_{{y}} {u}_{{ }{x}}-{u} {u}_{{x}{y}} )]\\

\end{array}
\en
\bn
\begin{array}{ll}

\mathcal{Q}_4 =&\frac{1}{4t\sqrt{y}}(4ty {u}_y+4xy{u}_{xxy}+t{u}+x{u}_{xx}+2y{u}_{xy})\\
\Phi_t^4=
& \frac{1}{16 t \sqrt{y}}[4 t y {u}_{{x}}+\left(2 y {u}_{{x}{y}}+x {u}_{{x}{x}}+4 x y {u}_{{x}{x}{y}}\right) {u}_{{x}}-{u} \left(4 t y {u}_{{ }{y}}+{u}_{{ }{x}{x}}+6 y {u}_{{x}{x}t{y}}+x {u}_{{x}{x}{x}}+4 x y {u}_{{x}{x}{x}{y}}\right)] \\

\mathcal{Q}_5 =&\frac{u_{xxx}}{y^{2/3}}\\
 \Phi_t^6=& \frac{1}{4} \left({u}_{{x}} {u}_{{x}{x}{y}}-{u} {u}_{{x}{x}{x}{y}}\right) \\

 \end {array}
 \en
 \bn
 \begin{array}{ll}

\end{array}
\en

\section{Wave equations on Kerr spacetime}
\subsection{Introduction}
In 1963, R. P. Kerr proposed a metric that describes a massive rotating object. Since then, a huge number of papers about the structure and astrophysical applications of this spacetime appeared.
Several investigations in the literature have been aimed at that spacetime. Exacts symmetries of the Kerr spacetime are given in \cite{Kramer}. In \cite{appro}, the authors investigated on the approximate symmetries on Kerr spacetime and find the rescaling factor for the energy. In this section, we analyse the symmetry structure of wave equation on Kerr spacetime. Noether approach and direct construction are used to find conserved densities on this spacetime.

The line element in this spacetime is given by \cite{kerr}

\bn
\d s^2=\frac{\Delta}{\rho^2}\left[\d t -k \sin ^2\t \d \phi\right]^2-\frac{\sin^2\t}{\rho^2}[(r^2+k^2)\d \phi -k\d t]^2-\frac{\rho^2	}{\Delta}\d r^2 -\rho^2
\d \theta^2\en
where $\Delta=r^2-2Mr+k^2$ and $\rho^2=r^2 +a^2 \cos^2\t$. $M$ and $k$ represent the mass and the rotation parameter, respectively. The angular momentum of the object is $J=Mk$. The Gordon type  equation is given by \eqref{waveq}

\bn
\begin{array}{ll}
&0=\frac{1}{(2Mr-k^2-r^2)}\left[4\left(Mr-\frac{k^2}{2}-\frac{r^2}{2}\right)^2 \sin ^2\t u_{r,r} -2\left(Mr-\frac{k^2}{2}-\frac{r^2}{2}\right)\sin ^2\t u_{\t,\t} \right.\\&+2\sin ^2\t\left(k^2\left(Mr-\frac{k^2}{2}-\frac{r^2}{2}\right)\cos ^2 \t -Mk^2r -\frac{k^2r^2}{2}-\frac{r^4}{2} \right)u_{t,t}+(k^2\cos^2 \t -2Mr+r^2)u_{\phi,\phi}- \\&\left.4 \sin \t (Mu_{t\phi}kr\sin \t -(\sin \t(M-r)u_r-\frac{1}{2}u_\t\cos \t)\left(Mr-\frac{k^2}{2}+\frac{r^2}{2}\right)\right]-k(u)(r^2+k^2\cos ^2 \t)\sin ^2 \t  \label{wave10}

\end{array}
\en

The corresponding Lagrangian is

\bn
\begin{array}{ll}
L=&\frac {1}{2\,Mr-{k}^{2}-{r}^{2}}[ \left( {k}^{2}\sin \t  \left( \cos \left(
\theta \right)  \right) ^{2} \left( Mr-\frac{1}{2}{k}^{2}-\frac{1}{2}{r}^{2}
 \right) -M{k}^{2}r\sin \t -\frac{1}{2}{k}^{2}{r}^{2}\sin
 \t -\frac{1}{2}\sin \t {r}^{4}
 \right) {u_{{t}}}^{2}]\\&-2{\frac {kMr\sin
 \t u_{{t}}u_{{\phi}}}{2Mr-{k}^{2}-{r}^{2}}}+\frac{1}{2}
\,\sin \t  \left( 2\,Mr-{k}^{2}-{r}^{2} \right) {u_
{{r}}}^{2}-\frac{1}{2}\sin \t {u_{{\theta}}}^{2}+\\&\frac{1}{2}{
\frac { \left( {k}^{2} \left( \cos \t  \right) ^{2}
-2Mr+{r}^{2} \right) {u_{{\phi}}}^{2}}{\sin \t
 \left( 2Mr-{k}^{2}-{r}^{2} \right) }}-h(u)\sin \theta (r^2+k^2\cos \theta)\label{wave11}
\end{array}
\en
where $h(u)=\int k(u) \d u$
\subsection{Symmetries of the waves equation-the Noether approach}

Many of the calculation have been left out as they are tedious. We classify the cases that yield strict Noether symmetries  (zero gauge) of \eqref{wave10}

The principal Noether algebra is for the case $h(u)=0$ in \eqref{wave11}

\bn
\begin{array}{ll}
X_1=&\p_u\\
X_3=&\p_t\\
X_2=&\p_{\phi}
\end{array}
\en
The associated conserved vectors are

\bn\begin{array}{ll}
\Phi_1^{\phi}= & {\frac {-2kMru_{{t}}+2M \left( \cos \left(\theta \right)  \right) ^{2}kru_{{t}}-2Mru_{{\phi}}+ \left( \cos \left( \theta\right)  \right) ^{2}{k}^{2}u_{{\phi}}+{r}^{2}u_{{\phi}}}{\sin \left( \theta\right)  \left( 2Mr-{k}^{2}-{r}^{2} \right) }} \\
\Phi_1^{\theta}=& \sin \t u_{{\theta}} \\
\Phi_1^{r}=&  \left( 2Mr-{k}^{2}-{r}^{2} \right) u_{{r}}\sin \left(\theta \right)\\
\Phi_1^{t}=& -{\frac {\sin \t  \left( 2M \left( \cos \t  \right) ^{2}{k}^{2}ru_{{t}}- \left( \cos \t  \right) ^{2}{k}^{4}u_{{t}}- \left( \cos \t  \right) ^{2}{k}^{2}{r}^{2}u_{{t}}-2 M{k}^{2}ru_{{t}}-{k}^{2}{r}^{2}u_{{t}}-u_{{t}}{r}^{4}-2kMru_{{\phi}} \right) }{2Mr-{k}^{2}-{r}^{2}}} \\

\Phi_2^{t}=& {\frac {\sin \t  \left( 2\,M \left( \cos \t  \right) ^{2}{k}^{2}ru_{{t}}- \left( \cos \t  \right) ^{2}{k}^{4}u_{{t}}- \left( \cos \t  \right) ^{2}{k}^{2}{r}^{2}u_{{t}}
\mbox{}-2\,M{k}^{2}ru_{{t}}-{k}^{2}{r}^{2}u_{{t}}-u_{{t}}{r}^{4}-2\,kMru_{{\phi}} \right)
\mbox{}u_{{\phi}}}{2\,Mr-{k}^{2}-{r}^{2}}} \\

\Phi_3^{t}=& \frac {1}{2{\sin \t  \left( 2\,Mr-{k}^{2}-{r}^{2} \right) }}\left[-4\,{M}^{2}{r}^{2}{u_{{r}}}^{2}-4\,M \left( \cos \t  \right) ^{2}{k}^{2}r{u_{{r}}}^{2}-2\,M \left( \cos \t  \right) ^{4}{k}^{2}r{u_{{t}}}^{2}
\mbox{}+\right. \\& 4\,M \left( \cos \t  \right) ^{2}{k}^{2}r{u_{{t}}}^{2}-{k}^{2}{u_{{\theta}}}^{2}-{r}^{2}{u_{{\theta}}}^{2}-{r}^{4}{u_{{t}}}^{2}-{r}^{4}{u_{{r}}}^{2}-{k}^{4}{u_{{r}}}^{2}-{r}^{2}{u_{{\phi}}}^{2}
\mbox{}-2\,M{k}^{2}r{u_{{t}}}^{2}+\\& \left( \cos \t  \right) ^{4}{k}^{2}{r}^{2}{u_{{t}}}^{2}+4\,{M}^{2} \left( \cos \t  \right) ^{2}{r}^{2}{u_{{r}}}^{2}-4\,M \left( \cos \t  \right) ^{2}{r}^{3}{u_{{r}}}^{2}
\mbox{}+2\, \left( \cos \t  \right) ^{2}{k}^{2}{r}^{2}{u_{{r}}}^{2}-\\&2\,M \left( \cos \t  \right) ^{2}r{u_{{\theta}}}^{2}+4\,M{k}^{2}r{u_{{r}}}^{2}+2\,Mr{u_{{\theta}}}^{2}-2\,{k}^{2}{r}^{2}{u_{{r}}}^{2}
\mbox{}+4\,M{r}^{3}{u_{{r}}}^{2}+ \left( \cos \t  \right) ^{4}{k}^{4}{u_{{t}}}^{2}+ \\&\left( \cos \t  \right) ^{2}{k}^{4}{u_{{r}}}^{2}+ \left( \cos \t  \right) ^{2}{r}^{4}{u_{{t}}}^{2}
\mbox{}+ \left( \cos \t  \right) ^{2}{r}^{4}{u_{{r}}}^{2}+ \left( \cos \t  \right) ^{2}{k}^{2}{u_{{\theta}}}^{2}+ \left( \cos \t  \right) ^{2}{r}^{2}{u_{{\theta}}}^{2}-\\&\left. \left( \cos \t  \right) ^{2}{k}^{4}{u_{{t}}}^{2}
\mbox{}+2\,Mr{u_{{\phi}}}^{2}- \left( \cos \t  \right) ^{2}{k}^{2}{u_{{\phi}}}^{2}-{k}^{2}{r}^{2}{u_{{t}}}^{2} \right]\\
\end{array}
\en
\subsection{Symmetries of the waves equation-the multipliers approach}

Consider the wave equation \eqref{wave10} with $k(u)=0$. We have

\bn
\begin{array}{ll}
\frac{\delta}{\delta u}[ \mathcal{Q}&\frac{1}{(2Mr-k^2-r^2)}\left[4\left(Mr-\frac{k^2}{2}-\frac{r^2}{2}\right)^2 \sin ^2\t u_{rr} -2\left(Mr-\frac{k^2}{2}-\frac{r^2}{2}\right)\sin ^2\t u_{\t\t} \right.\\&+2\sin ^2\t\left(k^2\left(Mr-\frac{k^2}{2}-\frac{r^2}{2}\right)\cos ^2 \t -Mk^2r -\frac{k^2r^2}{2}-\frac{r^4}{2} \right)u_{t,t}+(k^2\cos^2 \t -2Mr+r^2)u_{\phi\phi}- \\&\left.4 \sin \t (Mu_{t\phi}kr\sin \t -(\sin \t(M-r)u_r-\frac{1}{2}u_\t\cos \t)\left(Mr-\frac{k^2}{2}+\frac{r^2}{2}\right)\right]]\\&=D_t\Phi^t+D_r\Phi^r+D_{\t}\Phi^{\t}+
D_{\phi}\Phi^{\phi}
\end{array}
\en

%
%
%

where $\mathcal{Q}=\mathcal{Q}(u_\phi ,u_t, u_{\phi,\phi},u_{t,t},u_{t,\phi},u_{\phi,\phi,t},u_{t,\phi,t},u_{\phi,\phi,\phi})$

After tedious calculations, we obtain a set of multipliers $\mathcal{Q}_i$ together with the conserved densities.

\bn
\begin{array}{ll}
\mathcal{Q}_1=&{u}_{t}\\

 \Phi_1^t=&\frac{1}{2 (k^2+r (r-2
   M)) \sin (\theta )}[\sin (\theta )^2 u_{\text{ }t} (2 k M r u_{\text{ }\phi
   }+((k^2+r (r-2 M)) \cos (\theta ) k^2+r (r^3+k^2 (2
   M+r))) u_{\text{ }t})-\\&u (2 k^2 \cos
   (\theta )^2+(k^2+r (r-2 M)) \sin (\theta ) u_{\text{ }\theta }
   \cos (\theta )+2 r (r-2 M)+\sin (\theta )^2 (u_{\text{ }rr} k^4+2 r^2
   u_{\text{ }rr} k^2-\\&4 M r u_{\text{ }rr} k^2-2 M r u_{\text{ }t\phi }
   k+(k^2+r (r-2 M)) u_{\text{ }\theta \theta }-2 (M-r)
   (k^2+r^2-2 M r) u_{\text{ }r}+r^4 u_{\text{ }rr}-\\&4 M r^3
   u_{\text{ }rr}+4 M^2 r^2 u_{\text{ }rr}))] \\

%
    \mathcal{Q}_2=&\text{1}\\

 \Phi_2^t=&\frac{1}{k^2+r (r-2 M)}[\sin (\theta ) \left(2 k M r u_{\text{ }\phi }+\left(\left(k^2+r (r-2
   M)\right) \cos (\theta ) k^2+r \left(r^3+k^2 (2 M+r)\right)\right) u_{\text{
   }t}\right)]\\


%

   \mathcal{Q}_3=&{u}_\phi\\
\Phi_3^t=&\frac{1}{2 (k^2+r (r-2
   M))}[\sin (\theta ) (2 k M r u_{\text{ }\phi }{}^2+((k^2+r (r-2
   M)) \cos (\theta ) k^2+r (r^3+k^2 (2 M+r))) u_{\text{
   }t} u_{\text{ }\phi }-\\&u (2 k M r u_{\text{}\phi
   \phi }+((k^2+r (r-2 M)) \cos (\theta ) k^2+r (r^3+k^2 (2
   M+r))) u_{\text{ }t\phi }))] \\

   \mathcal{Q}_4=&{u}_{\phi\phi\phi}\\

\Phi_4^t=&\frac{1}{2 \left(k^2+r (r-2 M)\right)}[\sin (\theta ) (2 k M r u_{\text{}\phi } u_{\text{ }\phi \phi \phi
   }+((k^2+r (r-2 M)) \cos (\theta ) k^2+r (r^3+k^2 (2
   M+r))) u_{\text{ }t} u_{\text{ }\phi \phi \phi }-\\&u (2 k M r u_{\text{}\phi \phi \phi \phi }+((k^2+r (r-2
   M)) \cos (\theta ) k^2+r (r^3+k^2 (2 M+r))) u_{\text{
   }t\phi \phi \phi }))] \\

   \end{array}\en
   \bn
   \begin{array}{ll}

\mathcal{Q}_5=&{u}_{tt\phi}\\

\Phi_5^t=&\frac{1}{6
  (k^2+r (r-2 M)) \sin (\theta )}[-4 k^2 u_{\text{ }t\phi } \cos (\theta )^2+(k^2+r (r-2 M))
   \sin (\theta ) (2 \sin (\theta ) u_{\text{ }t\phi } u_{\text{ }tt}
   k^2+\\&2 \sin (\theta ) u_{\text{ }t} u_{\text{ }tt\phi } k^2-\sin (\theta )
   u_{\text{ }\phi } u_{\text{ }ttt} k^2-\sin (\theta ) u
   u_{\text{}ttt\phi } k^2+u_{\text{ }\theta \phi } u_{\text{ }t}-2
   u_{\text{ }\theta } u_{\text{ }t\phi }+u_{\text{ }\phi } u_{\text{
   }t\theta }-\\&2 u u_{\text{ }t\theta \phi }) \cos
   (\theta )+4 (2 M-r) r u_{\text{ }t\phi }+\sin (\theta )^2 (u_{\text{
   }rr\phi } u_{\text{ }t} k^4-2 u_{\text{ }rr} u_{\text{ }t\phi }
   k^4+u_{\text{}\phi } u_{\text{ }trr} k^4-2 u u_{\text{
   }trr\phi } k^4+\\&2 r^2 u_{\text{ }rr\phi } u_{\text{ }t} k^2-4 M r u_{\text{
   }rr\phi } u_{\text{ }t} k^2-2 u_{\text{ }\theta \theta } u_{\text{ }t\phi }
   k^2+4 M u_{\text{ }r} u_{\text{ }t\phi } k^2-4 r u_{\text{ }r} u_{\text{
   }t\phi } k^2-4 r^2 u_{\text{ }rr} u_{\text{ }t\phi } k^2+\\&8 M r u_{\text{
   }rr} u_{\text{ }t\phi } k^2+u_{\text{ }\phi } u_{\text{ }t\theta \theta }
   k^2-2 u u_{\text{ }t\theta \theta \phi } k^2-2 M
   u_{\text{ }\phi } u_{\text{ }tr} k^2+2 r u_{\text{ }\phi } u_{\text{ }tr}
   k^2+4 M u u_{\text{ }tr\phi } k^2-4 r u
   u_{\text{ }tr\phi } k^2\\&+2 r^2 u_{\text{}\phi } u_{\text{ }trr} k^2-4 M r
   u_{\text{ }\phi } u_{\text{ }trr} k^2-4 r^2 u u_{\text{
   }trr\phi } k^2+8 M r u u_{\text{ }trr\phi } k^2+2 r^2
   u_{\text{ }t\phi } u_{\text{ }tt} k^2+4 M r u_{\text{ }t\phi } u_{\text{
   }tt} k^2\\&+2 r^2 u_{\text{ }t} u_{\text{ }tt\phi } k^2+4 M r u_{\text{ }t}
   u_{\text{ }tt\phi } k^2-r^2 u_{\text{ }\phi } u_{\text{ }ttt} k^2-2 M r
   u_{\text{ }\phi } u_{\text{ }ttt} k^2-r^2 u u_{\text{
   }ttt\phi } k^2-2 M r u u_{\text{ }ttt\phi } k^2+\\& 8 M r
   u_{\text{ }t\phi }{}^2 k-4 M r u_{\text{ }t} u_{\text{ }t\phi \phi } k+2 M
   r u_{\text{ }\phi } u_{\text{ }tt\phi } k+2 M r u
   u_{\text{ }tt\phi \phi } k+\left(k^2+r (r-2 M)\right) u_{\text{ }\theta
   \theta \phi } u_{\text{ }t}-\\&2 (M-r) \left(k^2+r^2-2 M r\right) u_{\text{
   }r\phi } u_{\text{ }t}+r^4 u_{\text{ }rr\phi } u_{\text{ }t}-4 M r^3
   u_{\text{ }rr\phi } u_{\text{ }t}+4 M^2 r^2 u_{\text{ }rr\phi } u_{\text{
   }t}-2 r^2 u_{\text{ }\theta \theta } u_{\text{ }t\phi }+\\&4 M r u_{\text{
   }\theta \theta } u_{\text{}t\phi }-4 r^3 u_{\text{ }r} u_{\text{, }t\phi
   }+12 M r^2 u_{\text{ }r} u_{\text{ }t\phi }-8 M^2 r u_{\text{ }r}
   u_{\text{ }t\phi }-2 r^4 u_{\text{ }rr} u_{\text{ }t\phi }+8 M r^3
   u_{\text{ }rr} u_{\text{ }t\phi }-\\&8 M^2 r^2 u_{\text{ }rr} u_{\text{
   }t\phi }+r^2 u_{\text{ }\phi } u_{\text{ }t\theta \theta }-2 M r u_{\text{
   }\phi } u_{\text{ }t\theta \theta }-2 r^2 u u_{\text{
   }t\theta \theta \phi }+4 M r u u_{\text{}t\theta \theta
   \phi }+2 r^3 u_{\text{ }\phi } u_{\text{}tr}-6 M r^2 u_{\text{ }\phi }
   u_{\text{ }tr}\\&+4 M^2 r u_{\text{ }\phi } u_{\text{ }tr}-4 r^3 u u_{\text{ }tr\phi }+12 M r^2 u u_{\text{ }tr\phi
   }-8 M^2 r u u_{\text{ }tr\phi }+r^4 u_{\text{ }\phi }
   u_{\text{ }trr}-4 M r^3 u_{\text{ }\phi } u_{\text{ }trr}+\\&4 M^2 r^2
   u_{\text{ }\phi } u_{\text{ }trr}-2 r^4 u u_{\text{
   }trr\phi }+8 M r^3 u u_{\text{ }trr\phi }-8 M^2 r^2
   u u_{\text{ }trr\phi }+2 r^4 u_{\text{ }t\phi }
   u_{\text{ }tt}+2 r^4 u_{\text{ }t} u_{\text{ }tt\phi }-r^4 u_{\text{}\phi
   } u_{\text{ }ttt}\\&-r^4 u u_{\text{ }ttt\phi })]\\
   \end{array}\en
  \bn
  \begin{array}{ll}

\mathcal{Q}_6=&{u}_{t\phi\phi}\\

\Phi_6^t=&-\frac{1}{6 (k^2+r (r-2
   M)) \sin (\theta )}[u_{\text{ }\phi \phi } (2 k^2 \cos (\theta )^2-(k^2+r (r-2
   M)) \sin (\theta ) (k^2 \sin (\theta ) u_{\text{ }tt}-u_{\text{
   }\theta }) \cos (\theta )+\\&2 r (r-2 M)+\sin (\theta )^2 (u_{\text{
   }rr} k^4+2 r^2 u_{\text{ }rr} k^2-4 M r u_{\text{ }rr} k^2-r^2 u_{\text{
   }tt} k^2-2 M r u_{\text{ }tt} k^2-4 M r u_{\text{ }t\phi } k+\\&(k^2+r
   (r-2 M)) u_{\text{ }\theta \theta }-2 (M-r) (k^2+r^2-2 M r)
   u_{\text{ }r}+r^4 u_{\text{ }rr}-4 M r^3 u_{\text{ }rr}+4 M^2 r^2
   u_{\text{ }rr}-r^4 u_{\text{ }tt}))\\&+\sin (\theta ) (\sin
   (\theta ) (-3 r (r^3+k^2 (2 M+r)) u_{\text{ }t} u_{\text{
   }t\phi \phi }-u_{\text{}\phi } (u_{\text{ }rr\phi } k^4+2 r^2
   u_{\text{ }rr\phi } k^2-4 M r u_{\text{}rr\phi } k^2-r^2 u_{\text{ }tt\phi
   } k^2\\&-2 M r u_{\text{ }tt\phi } k^2+2 M r u_{\text{ }t\phi \phi }
   k+(k^2+r (r-2 M)) u_{\text{ }\theta \theta \phi }-2 (M-r)
   (k^2+r^2-2 M r) u_{\text{ }r\phi }+r^4 u_{\text{}rr\phi }\\&-4 M
   r^3 u_{\text{ }rr\phi }+4 M^2 r^2 u_{\text{ }rr\phi }-r^4 u_{\text{ }tt\phi
   })+u (u_{\text{}rr\phi \phi } k^4+2 r^2
   u_{\text{ }rr\phi \phi } k^2-4 M r u_{\text{ }rr\phi \phi } k^2+2 r^2
   u_{\text{ }tt\phi \phi } k^2+\\&4 M r u_{\text{ }tt\phi \phi } k^2+2 M r
   u_{\text{ }t\phi \phi \phi } k+(k^2+r (r-2 M)) u_{\text{ }\theta
   \theta \phi \phi }-2 (M-r) (k^2+r^2-2 M r) u_{\text{ }r\phi \phi
   }+r^4 u_{\text{}rr\phi \phi }\\&-4 M r^3 u_{\text{ }rr\phi \phi }+4 M^2 r^2
   u_{\text{}rr\phi \phi }+2 r^4 u_{\text{ }tt\phi \phi
   }))+(k^2+r (r-2 M)) \cos (\theta ) (-3 \sin
   (\theta ) u_{\text{ }t} u_{\text{ }t\phi \phi } k^2+\\&u_{\text{ }\phi }
  (k^2 \sin (\theta ) u_{\text{}tt\phi }-u_{\text{ }\theta \phi
   })+u (2 \sin (\theta ) u_{\text{}tt\phi \phi }
   k^2+u_{\text{ }\theta \phi \phi })))] \\

    \end{array}\en
  \bn
  \begin{array}{ll}
&+12 M r^2 u_{\text{ }r} u_{\text{
   }t\phi }-8 M^2 r u_{\text{ }r} u_{\text{ }t\phi }-2 r^4 u_{\text{ }rr}
   u_{\text{ }t\phi }+8 M r^3 u_{\text{ }rr} u_{\text{ }t\phi }-8 M^2 r^2
   u_{\text{ }rr} u_{\text{}t\phi }+r^2 u_{\text{ }\phi } u_{\text{}t\theta
   \theta }-\\&2 M r u_{\text{}\phi } u_{\text{ }t\theta \theta }-2 r^2
   u u_{\text{}t\theta \theta \phi }+4 M r u u_{\text{ }t\theta \theta \phi }+2 r^3 u_{\text{ }\phi } u_{\text{
   }tr}-6 M r^2 u_{\text{ }\phi } u_{\text{ }tr}+4 M^2 r u_{\text{}\phi }
   u_{\text{}tr}-4 r^3 u u_{\text{}tr\phi }+\\&12 M r^2
   u u_{\text{ }tr\phi }-8 M^2 r u
   u_{\text{ }tr\phi }+r^4 u_{\text{ }\phi } u_{\text{ }trr}-4 M r^3
   u_{\text{ }\phi } u_{\text{ }trr}+4 M^2 r^2 u_{\text{ }\phi } u_{\text{
   }trr}-2 r^4 u u_{\text{ }trr\phi }+\\&8 M r^3 u
    u_{\text{ }trr\phi }-8 M^2 r^2 u u_{\text{
   }trr\phi }+2 r^4 u_{\text{}t\phi } u_{\text{ }tt}-r^4 u_{\text{ }t}
   u_{\text{ }tt\phi }-r^4 u_{\text{ }\phi } u_{\text{ }ttt}+2 r^4
   u u_{\text{ }ttt\phi })] \\

\end{array}
\en


\begin{thebibliography}{99}

\bibitem{221} H Azad and M T Mustafa, Symmetry analysis of waves equation on sphere,\textit{Journal of Mathematical Analysis and Applications}, 333, 1180,  2007.

\bibitem{113}{ G Bluman and Z Yang},
{Some recent developments in finding systematically conservation laws and nonlocal symmetries for partial differential equations}, \textit{Lecture Notes in Applied and Computational Mechanics}, {73}, {2014}.


\bibitem{mine}J J H Bashingwa, A H Kara, Ashfaque H. Bokhari, R A Mousa, F D Zaman, Symmetry and conservation law of some anti-self-duality (ASD) manifolds, \textit{Pramana Journal of Physics} (to appear), (2016).

\bibitem{d} M Dunajski, An interpolating dispersionless integrable system, \emph{J. Phys. A: Math. Theor}. 41, 315202, 2008.

\bibitem{f} K Fukaya, Anti-self-dual Equation on 4-manifolds with Degenerate Metric, \emph{Geometric \& Functional Analysis}, 8(3), 466, 1998.


\bibitem{appro} I Hussain, F M Mahomed, and A Qadir, Approximate Noether symmetries of the geodesic equations for the charged-kerr spacetime and rescaling of the energy, \textit{General Relativity and gravitation}, 411.10, 2399, 2009.

\bibitem{Ibra} N H Ibragimov (eda),  \textit{CRC handbook of Lie group analysis of differential equations}, {CRC} Press, Boca Raton, Vol.1,  1996.
%
\bibitem{222} S Jamal, An analysis of the symmetries and conservation laws of some classes of nonlinear waves equations in curved spacetime geometry, \textit{A thesis submited to the Faculty of Science, University of the Witwatersrand, Johannesburg } 2013.
    




\bibitem{115}{A Jhangeer and S Sharif}, {Conserved quantities and group classification of wave equation on hyperbolic space}, \textit{Communications in Nonlinear Science and Numerical Simulation}, {18}, {236}, {2013}.


\bibitem{kerr} R P Kerr, Gravitational Field of a spinning Mass as an Example of Algebraically Special Metrics. \textit{Physical Review letters}, {11}, 237, 1963

\bibitem{Kramer} D Kramer D, H Stephani, A A H MacCullum and E Herlt, \textit{Exact Aolutions of Einstein Field Equations}, Cambridge University Press, Cambridge, 1980.

\bibitem{m} A A Malykh and M B Sheftel, General heavenly equation governs anti-self-dual gravity, \emph{J. Phys. A.: Math. Theor.} 44,  155201, 2011.

\bibitem{N11} E. Noether, Invariante Variationsprobleme, {\it Nachrichten der Akademie
der Wissenschaften in G\"ottingen, Mathematisch-Physikalische
Klasse}, {\bf 2}, 235-257, 1918 (English translation in {\it
Transport Theory and Statistical Physics}, {\bf 1}(3), 186-207, 1971.

\bibitem{p} J F Plebanski, Some solutions of complex Einstein equations, \emph{J. Math. Phys}. 16, 2395, 1975.







\bibitem{09} P Olver, {\it Application of Lie groups to differential equations }
Springer-Verlag, New York, 1986.


\bibitem{S10} H Stephani, { Differential Equations: their solution using symmetries}
Cambridge University Press, Cambridge, 1989.























\bibitem{111} Y Wang, Y and L Wei, {Auxiliary Lagrangian and conservation laws for a wave equation incorporating dissipation},
\textit{Communications in Theoretical Physics}, {63}, {481}, {2015},.


\bibitem{112}{Y Yun and C Temuer},
{Classical and nonclassical symmetry classifications of nonlinear wave equation with dissipation}, \textit{Applied Mathematics and Mechanics (English Edition)}, {36}, {365}, {2015}.


























\end{thebibliography}
\end{document}